\documentclass[10pt,twocolumn,showpacs,superscriptaddress]{revtex4}
\usepackage[latin9]{inputenc}
\usepackage{amsmath}
\usepackage{amssymb}
\usepackage{graphicx}

\makeatletter
\@ifundefined{textcolor}{}
{%
 \definecolor{BLACK}{gray}{0}
 \definecolor{WHITE}{gray}{1}
 \definecolor{RED}{rgb}{1,0,0}
 \definecolor{GREEN}{rgb}{0,1,0}
 \definecolor{BLUE}{rgb}{0,0,1}
 \definecolor{CYAN}{cmyk}{1,0,0,0}
 \definecolor{MAGENTA}{cmyk}{0,1,0,0}
 \definecolor{YELLOW}{cmyk}{0,0,1,0}
}

\pdfoutput=1

\makeatother

\begin{document}

\title{Radiation Hardness of High-Q Silicon Nitride Microresonators for
Space Compatible Integrated Optics}

\author{Victor Brasch}

\affiliation{École Polytechnique Fédérale de Lausanne, CH-1015, Switzerland}

\author{Qun-Feng Chen}

\author{Stephan Schiller}

\email{Corresponding author: step.schiller@hhu.de}

\affiliation{Institut für Experimentalphysik, Heinrich-Heine-Universität Düsseldorf,
DE-40225, Germany}

\author{Tobias J. Kippenberg}

\email{Corresponding author: tobias.kippenberg@epfl.ch}

\affiliation{École Polytechnique Fédérale de Lausanne, CH-1015, Switzerland}
\begin{abstract}
Integrated optics has distinct advantages for applications in space
because it integrates many elements onto a monolithic, robust chip.
As the development of different building blocks for integrated optics
advances, it is of interest to answer the important question of their
resistance with respect to ionizing radiation. Here we investigate
effects of proton radiation on high-Q ($\mathcal{O}(10$$^{6})$)
silicon nitride microresonators formed by a waveguide ring. We show
that the irradiation with high-energy protons has no lasting effect
on the linear optical losses of the microresonators.
\end{abstract}
\maketitle
Optics has been an important part of space systems ever since the
early years of space exploration. While imaging optics were part of
some of the first instruments in space, the field of optics keeps
expanding and in particular the field of integrated optics shows great
potential for multiple applications with its distinct advantages of
small size, robustness against vibrations and high degree of integration.
The effect of ionizing radiation in space has been tested in many
studies for discrete optics\,\cite{Berghmans2008,Shetter1979,Appourchaux1993,Naletto2003,Chen2013}
as well as fiber optics\,\cite{Berghmans2008,Wall1975,Ott2002,Henschel2001,Lezius2012}.
From these studies it is well known that radiation can change the
optical properties of glasses which can lead to increased losses.
However, this important aspect of space compatibility has not been
investigated extensively for integrated optics\,\cite{Henschel1993,Passaro:02}.
Here we study the radiation hardness of waveguide microresonators
made from silicon nitride (SiN)\,\cite{Stutius1977,Moss2013} with
quality factors (Q-factors) around $10^{6}$. These microresonators
can find applications as optical frequency comb generators and optical
filters\,\cite{Levy2010,Herr2012,Barwicz2004}. The optical frequency
combs generated inside the resonators have frequency spacings of around
10\,GHz to 1\,THz and span several hundred nm\,\cite{Levy2010,Herr2012,Ferdous2011}.
Frequency combs can be crucial parts of future space missions and
applications. Examples for missions in the field of fundamental physics
that would benefit of such a device are the missions ``Space Optical
Clock'' (SOC)\,\cite{Schiller2012} on the ISS, the ``Einstein
Gravity Explorer'' (EGE)\,\cite{Schiller2009}, and the ``Space
Time Explorer and Quantum Test of the Equivalence Principle'' (STE-QUEST)\,\cite{Stequest2013,Chen2013}.
In the first and second mission, a frequency comb is required to convert
the laser radiation of an optical clock into microwave radiation that
is transmitted to ground where its frequency is measured. In the STE-QUEST
mission, a microwave-optical local oscillator has been proposed as
part of a microwave cold Cs atomic clock, which provides, again via
conversion by a frequency comb, an ultrastable 9\,GHz microwave for
interrogation of the atomic clock. 

The generation of microresonator frequency combs relies on the Kerr
nonlinearity of the silicon nitride resonator. The effect of the nonlinearity
is greatly enhanced by the high Q-factor of the resonator, the threshold
for the parametric oscillation scales as $1/Q^{2}$\,\cite{Kippenberg2004,Herr2012}.
Therefore the Q-factor of a microresonator is an important parameter
for these applications. The Q-factor is limited by the optical losses
of the waveguide ring resonator which are caused by absorption as
well as scattering. Therefore, a change in Q-factor directly relates
to changed losses of the waveguide.

One of the key requirements for space equipment is radiation resistance.
Satellites on orbits that cross the Van Allen belt are exposed strongly
to electron and proton radiation. Components of instruments that are
to be flown on such orbits must therefore be tested beforehand with
respect to their radiation resistance. The mentioned missions EGE
and STE-QUEST rely on such orbits. In this work, we perform a first
investigation of the influence of proton radiation on the Q-factor
as a key property of microresonators.

\begin{figure*}
\includegraphics[width=1.75\columnwidth]{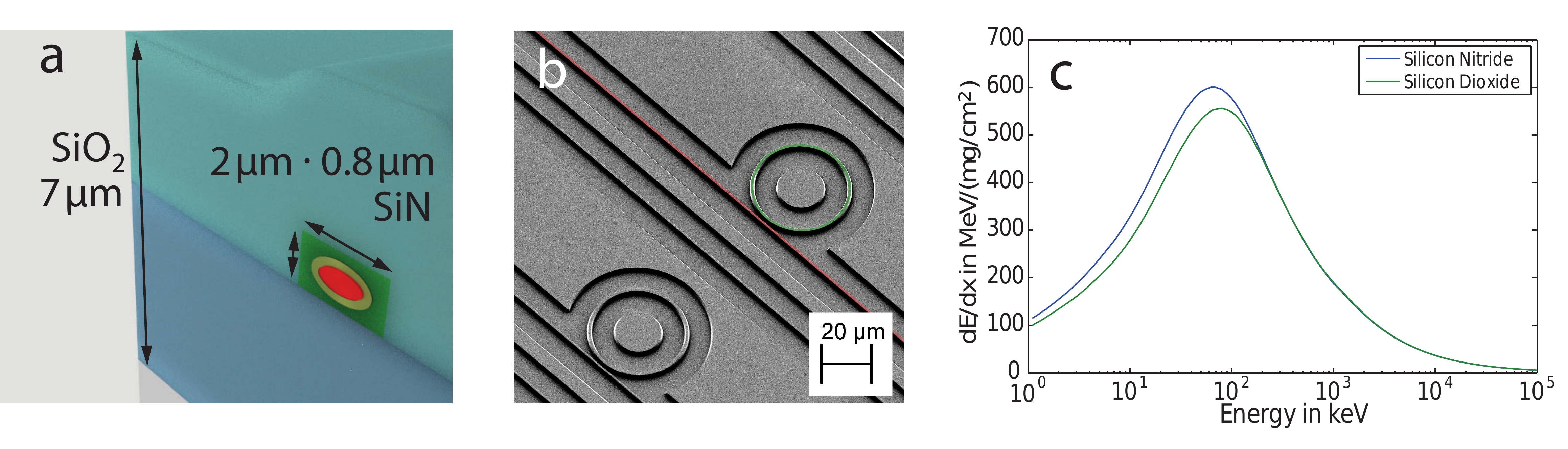}

\caption{\label{fig:energyLossFunc}(a) Schematic cross-section of the system
tested. The silicon nitride waveguide is shown in green with an approximated
intensity distribution for a fundamental mode around 1550\,nm inside
(red). The silicon dioxide cladding above and below differ slightly
as they have been deposited or grown using different methods. (b)
Colored electron microscopy image of SiN microresonators (not the
ones tested and with smaller radius) without the glass cladding. The
resonator waveguide is colored green while the bus waveguide is colored
red. (c) Energy loss function for protons with energies ranging from
1\,keV to 100\,MeV in silicon nitride and silicon dioxide calculated
with SRIM\,\cite{Ziegler2013}.}

\end{figure*}

The microresonators under test are waveguide ring resonators. A SiN
waveguide with dimensions 0.8\,$\mu$m by 2\,$\mu$m and with a
refractive index of 1.98 confines the light and is embedded into a
silicon dioxide (SiO$_{2}$) cladding with refractive index of approximately
1.45 (Fig.\,\ref{fig:energyLossFunc}(a)). From simulations we know
that at 1550\,nm about 80\,\% of the power travels inside the SiN and 20\,\%
in the SiO$_{2}$. To fabricate the ring microresonators, standard
p-doped (boron doping, resistivity of 10 to 20\,$\Omega$cm) 100\,mm
silicon wafers are oxidized in a thermal wet oxidation process in
order to grow a film of 4\,$\mu$m SiO$_{2}$. On top, 800\,nm of
silicon nitride (Si$_{3}$N$_{4}$) is deposited as nearly stoichiometric
high-stress thin film via low pressure chemical vapor deposition (LPCVD).
After patterning the silicon nitride using electron-beam lithography
and reactive-ion etching the 3\,$\mu$m thick SiO$_{2}$ top cladding
is deposited as low-temperature CVD oxide followed by a thermal anneal.
In a last step chips of 5\,mm by 5.5\,mm are separated. One chip
comprises multiple resonators. Each resonator is evanescently coupled
to one separate bus waveguide which allows to couple light in and
out of the resonator (Fig.\,\ref{fig:energyLossFunc}(b)). Further
details about the fabrication process can be found in \cite{Riemensberger2012}.

As outlined above, the Q-factor is one of the most important properties
of the resonator and the most likely to undergo significant changes
under irradiation\,\cite{Ott2002,MacPherson2004}. Therefore we selected
two chips from different wafers and two resonators on each chip and
characterized their Q-factors. The quality factor was measured by
determining the loaded linewidth ($\kappa/2\pi$ in Hz) of multiple
resonances between 1520 and 1580\,nm of the respective resonator.
The Q-factor relates to the linewidth as $Q=\omega/\kappa$ where
$\omega$ is the optical resonance frequency. The linewidths of the
resonances were measured by scanning an external cavity diode laser
with a linewidth of approximately 300\,kHz over the resonances. The
polarization does not have an effect on the linewidths of the resonance,
however, the contrast and the lineshape depend on the polarization.
Therefore the polarization of the laser was optimized by hand to yield
minimal transmission through the bus waveguide with the laser on resonance
and a lineshape with as little distortion from a Lorentzian shape
as possible. The laser scan was calibrated in frequency using a fiber
frequency comb which provides a frequency calibration marker approximately
every 60\,MHz with 1\,MHz precision\,\cite{Riemensberger2012,DelHaye2009}.
In most cases a good fit of the transmission curve could be obtained
with a simple Lorentzian lineshape function. For resonances that showed
a splitting due to the coupling of the co- and couter-propagating
modes, we fitted a model that takes this splitting into account\,\cite{Gorodetsky2000,Kippenberg2002}.
The average linewidths of the four measured resonators (resonator
I to IV) varied from 250\,MHz to 440\,MHz. 

\begin{figure}[t]
\includegraphics[width=0.95\columnwidth]{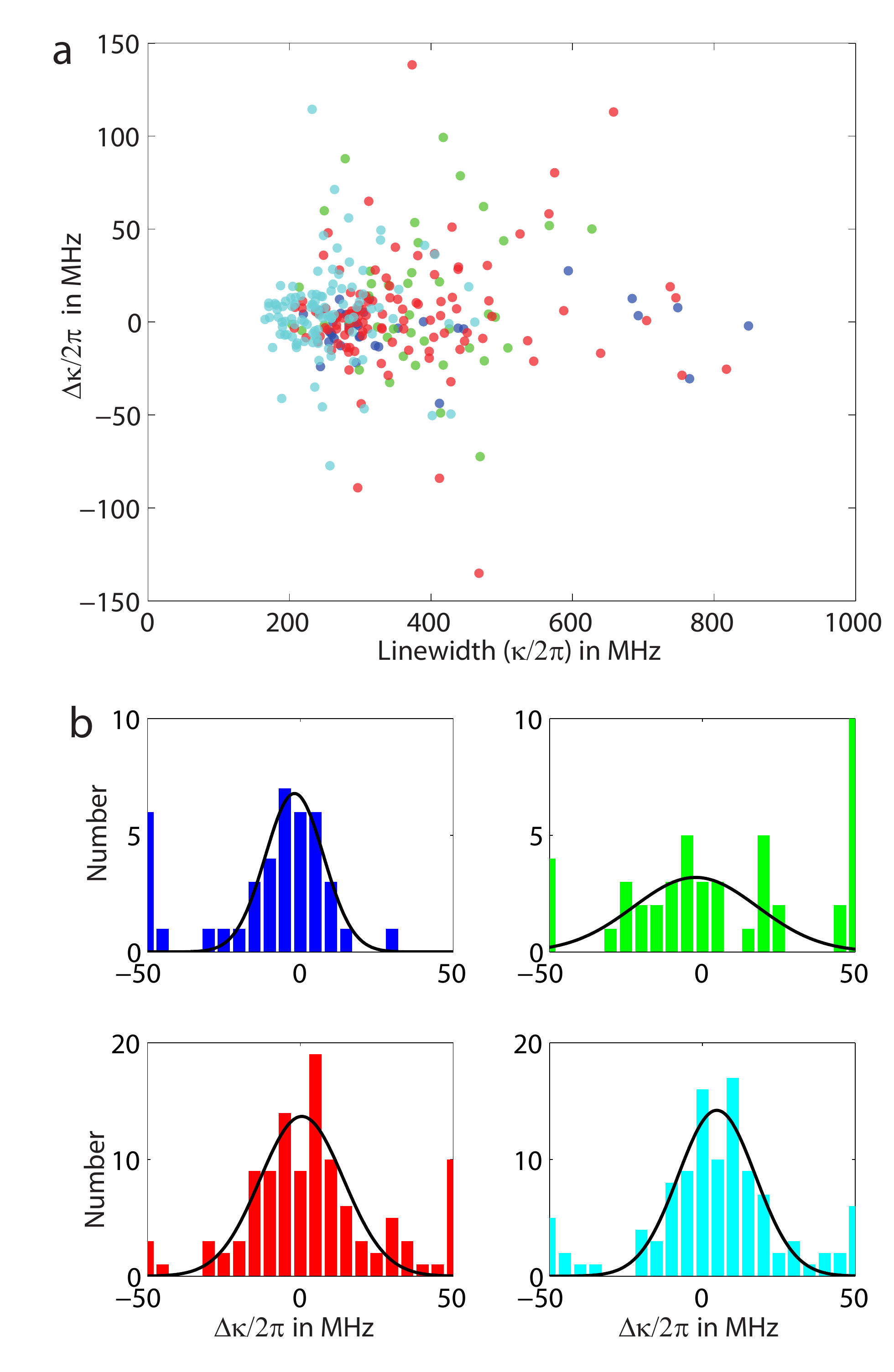}

\caption{\label{fig:results}(a) Change in total linewidth ($\Delta\kappa/2\text{\ensuremath{\pi}=\text{\ensuremath{\kappa}}}_{\text{after}}/2\pi-\text{\ensuremath{\kappa}}_{\text{before}}/2\pi$)
of the resonances of four resonators (dark blue, green, red, light
blue) plotted over averaged linewidth $(\kappa_{\text{after}}+\text{\ensuremath{\kappa}}_{\text{before}})/4\pi$.
Average linewidths for one resonator are 370, 440, 380 and 250\,MHz
for resonator I (dark blue), II (green), III (red) and IV (light blue)
respectively. (b) histograms of the change in linewidth for each resonator,
same dataset with same colors as in (a). The black lines represent
the fitted Gaussian distributions. The outermost bins of the histograms
on either side have been excluded for the fits. The values obtained
for the mean of the distribution as offset from 0 ($\langle\Delta\kappa_{f}\rangle/2\pi$)
are $-$1.9, $-$2.2, 0.5 and 4.7\,MHz for resonator I, II, III and
IV respectively. The standard deviation in the same order are 9.5,
20.3, 13.7 and 12.6\,MHz.}
\end{figure}

After the initial characterization the chips were sent for the irradiation.
After the irradiation was performed as described below the chips were
shipped back and the linewidths were measured again to check for changes.
For each resonance that was measured before the irradiation the polarization-dependent
shape of the resonance was again optimized to yield a minimal transmission
on resonance (for resonances of Chip 1 with resonators I and II, dark
blue and green data in Fig.\,\ref{fig:results} respectively) or
to obtain a best possible match with the shape of the resonance measured
before the irradiation (Chip 2 with resonator III and IV, red and
light blue data). Due to the shipments and preparation for the irradiation
as well as the extensive manual measurements the whole proceedure
streched over 12 weeks. Therefore, potentially short lasting irradiation
effects as they have been reported in fibers\,\cite{Griscom1993,Lezius2012}
are not well reflected in our measurements. 

Proton irradiation of the sample can have multiple effects which are
caused by the interaction with the material\,\cite{Berghmans2008}.
As the critical part of our chips is only around 8\,$\mu m$ thick,
most of the protons with energies above approximately 1\,MeV pass
through this part. The average energy deposited in the sample depends
on the energy loss function as it can be calculated for different
materials. In Fig.\,\ref{fig:energyLossFunc} the results of simulations
for the two relevant materials for this work, silicon nitride and
silicon dioxide, are shown.

\begin{figure}
\includegraphics[width=0.8\columnwidth]{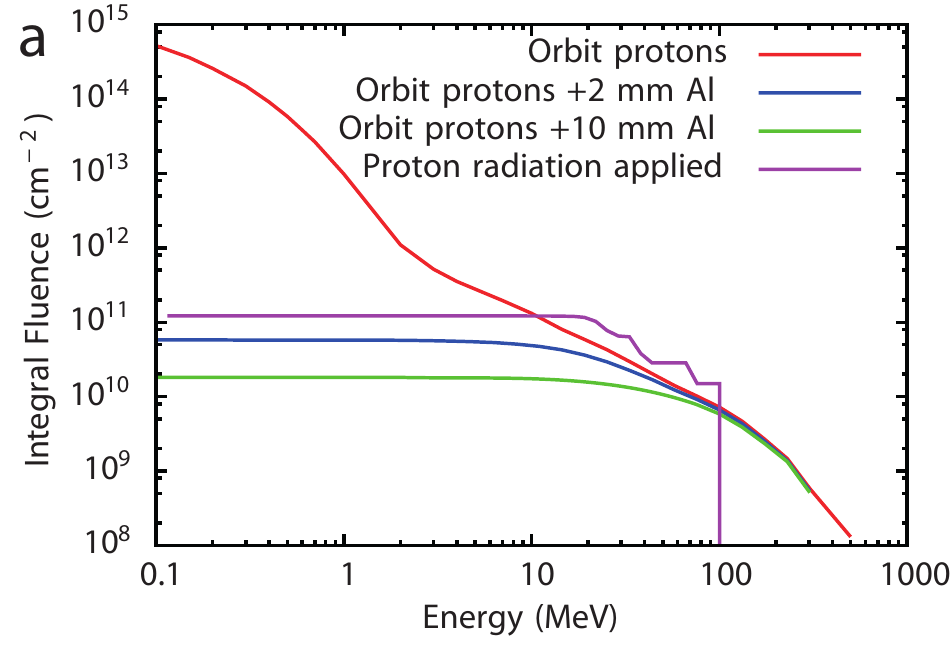}

\includegraphics[width=0.8\columnwidth]{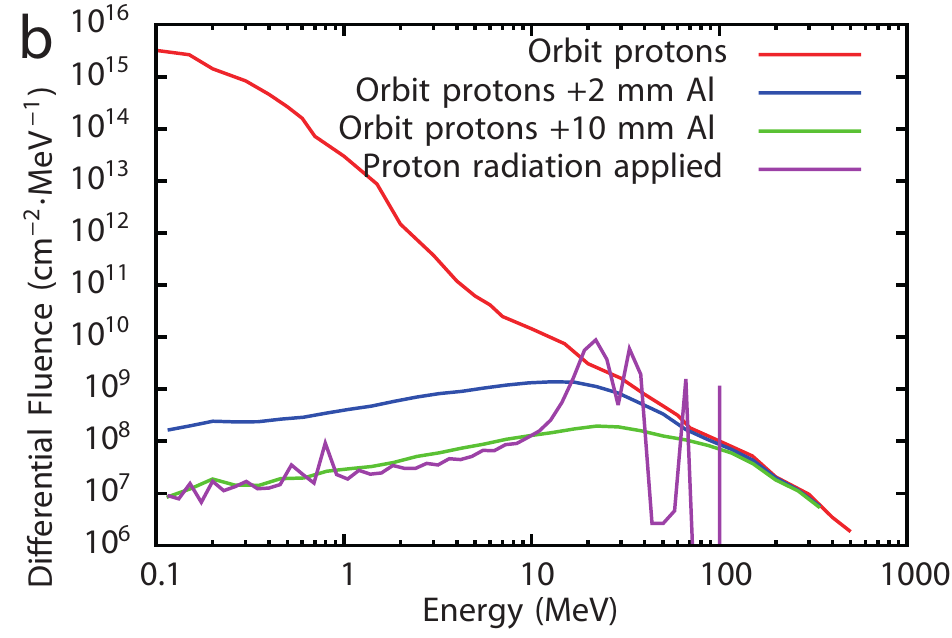}

\caption{\label{fig:radiationSpectrum} (a) Red, time-integrated and energy-integrated
(integral) proton fluence expected during the 5-year-long STE-QUEST
mission; blue, same, but behind a 2\,mm thick aluminum shield; green,
same, but behind a 10\,mm thick aluminum shield; purple, integral
fluence of the proton radiation applied in the test. (b) The corresponding
time-integrated differential fluences. The bin width at $E=99.7$\,MeV
in the plot (b) is 13\,MeV. ``Orbit protons'' refers to the proton
spectrum on the STE-QUEST orbin, with duration of 5\,years. ``Proton
radiation applied'' is the radiation impinging onto the tested microresonators.}

\end{figure}

The starting point of the irradiation study is the result of modeling
the proton spectrum experienced by a spacecraft on a specific orbit.
We assume here the highly elliptic orbit proposed for the STE-QUEST
mission. It is characterized by a 16-hour period, a perigee altitude
(above ground) varying between 800 and 2400\,km during the course
of the 5-year long mission, and a constant apogee altitude of approximately
51\,000\,km\,\cite{Stequest2013}. The time-averaged spectral (i.e.
energy-dependent or differential) fluence of the proton radiation,
and the corresponding time-averaged and energy-integrated (integral)
fluence of the protons, calculated in Ref.\,\cite{Stequest2012},
are shown in Fig.~\ref{fig:radiationSpectrum} (red line). The energy-integrated
fluence at energy $E$ is defined as the integral of the differential
fluence from infinite energy down to $E$. Note that the two quantities
have a similar energy dependence for this orbit. The quantities refer
to the radiation arriving on the satellite. 

When one considers the radiation reaching a particular component in
an instrument in the satellite, one must consider that this component
is shielded by other components of the satellite (satellite structure,
solar panels, housing of the instrument, additional specific shielding
layers) in a way that depends on the details of the spacecraft and
the instruments. The amount of shielding can be different in different
directions with respect to the spacecraft axes. The details are usually
not known with certainty a priori, since they are worked out during
the detailed planning of the mission, which only occurs after selection
of the mission by the space agency.

We therefore first consider as an example a shielding having an equivalent
thickness of 2\,mm aluminum (blue lines in Fig.\,\ref{fig:radiationSpectrum}).
Such a thickness is likely to be present even without installing additional
shielding. We see that the result is an extremely strong reduction
of the fluences at energies below 10\,MeV. The modified fluence was
computed using the software MULASSIS\,\cite{Lei2002,Mulassis}. In
this program, one layer of aluminum shielding is considered and the
modification of an input spectrum is then calculated, yielding an
output spectrum. 

We can assume the existence of 10\,mm aluminum shielding, which would
arise from structural shielding plus possibly an additional custom
housing enclosing the microresonator. The resulting fluences reaching
the component are shown in green in the plots. The increased shielding
thickness results in an additional factor 10 reduction in the fluences
below 10\,MeV. The fluences then have values that are well accessible
by an irradiation run produced by a proton accelerator. The task is
then to devise an irradiation protocol that models the predicted,
shielded space proton spectrum reasonably well.

The proton irradiation test was carried out at the Paul-Scherrer-Institut
(PSI) in Villigen, Switzerland. The maximum energy available at this
facility is 99.7\,MeV. This sets an upper limit for the energy of
the space proton spectrum that can be reproduced. Considering that
protons with energy lower than 20\,MeV would be effectively stopped
(i.e. absorbed) by a 2\,mm aluminum shield, the samples were irradiated
with protons with energies on the order of and higher than 20\,MeV.
The samples were irradiated with 4 energies, 18.3, 30.7, 61.6, and
99.7\,MeV. Except for the last one, these energies are the mean energies
after degradation of the proton beam by a copper plate (``degrader'')
of thickness 12.5, 11.5, and 7.5\,mm, respectively, inserted into
the proton beam. The fluences of the proton beam (upstream of the
respective degrader) were set to $6.000\times10^{10}$, $4.003\times10^{10}$,
$1.416\times10^{10}$, and $1.516\times10^{10}$ protons/${\rm cm}^{2}$,
respectively . 

The integral and differential fluence of the implemented irradiation
of the samples are shown in purple in Fig.\,\ref{fig:radiationSpectrum}.
They are the sum of four individual fluences, each of which corresponds
to the simulated fluence of a 99.7\,MeV proton beam degraded by the
respective copper degrader (if any).

In more detail, the differential fluence reproduces well the low-energy
region ($E<10$\,MeV) of the 10\,mm shielded space spectrum. In
the range 18 to 100\,MeV the spectrum consists of four peaks, rather
than a continuous spectrum, which are remainders of the energies of
the proton beam. We believe that this is of no consequence; in other
words, the potential damage done to the sample will likely not depend
on the details of the energy spectrum in the range $18-100$\,MeV.
The overall situation is that the implemented integral fluence exceeds
both the 10\,mm-shielded as well as the 2\,mm-shielded integral
space fluence at all energies $E<100$\,MeV. Therefore, the implemented
fluence is a conservative choice for a 5-year STE-QUEST mission duration
having a 10\,mm aluminum shielding. In addition, the integral fluence
down to $E=18$\,MeV is a factor 2 above the \textit{\emph{unshielded}}
STE-QUEST space spectrum (red line in the figure), also indicating
the conservative nature of our test irradiation protocol.

\begin{figure}[t]
\includegraphics[width=0.9\columnwidth]{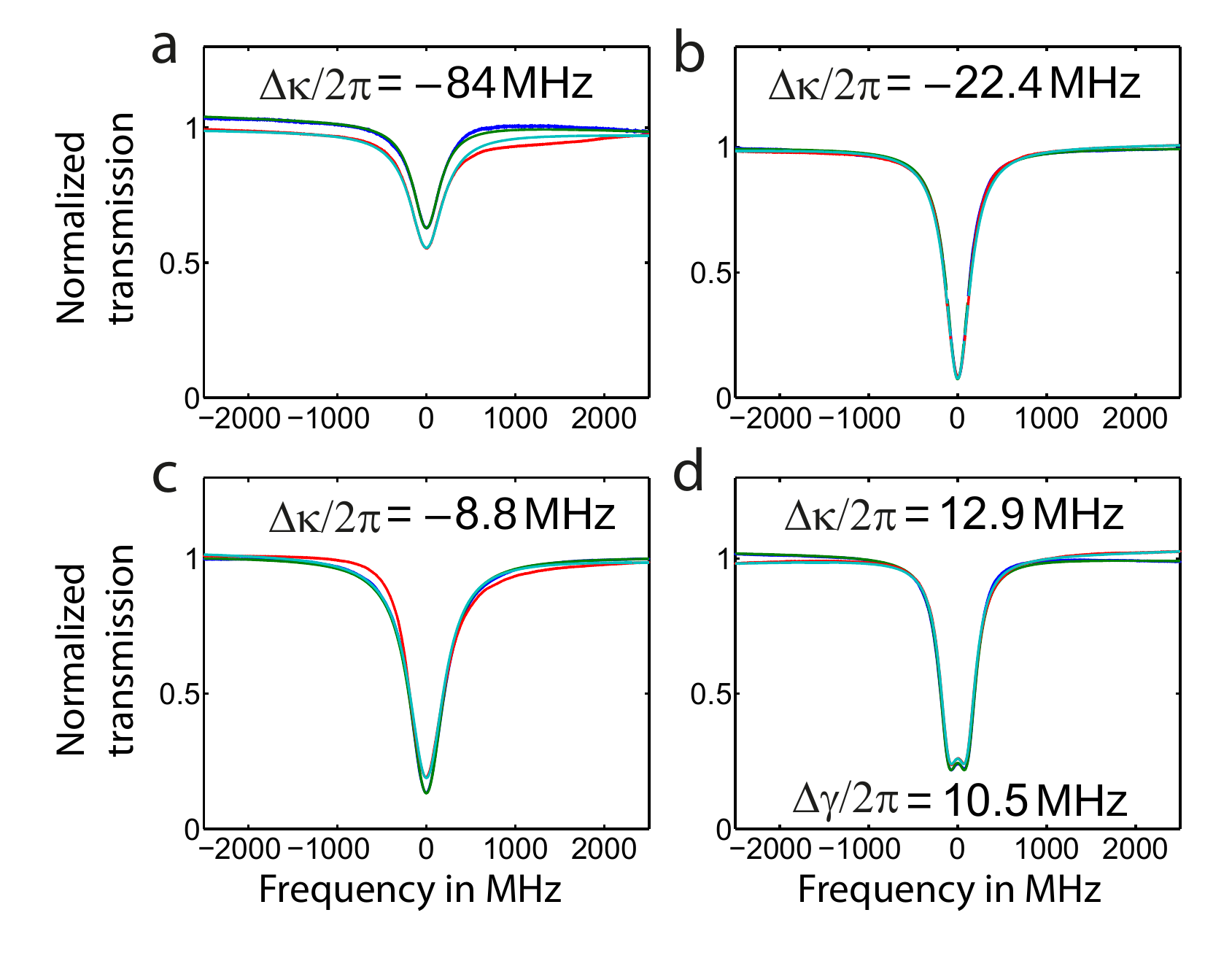}\caption{\label{fig:resonanceFits}Four measurements of total linewidths and
their fits from resonator III. Dark blue is the measurement before
radiation, red the measurement after radiation. Green and light blue
are the fits for pre and post radiation respectively. $\Delta\kappa/2\pi$
is the difference in fitted resonance widths as plotted in Fig.\,\ref{fig:results}.
$\Delta\gamma/2\pi$ is the difference in the fitted splitting of
the resonance in the two fits. These differences are caused by the
polarization dependency of the exact lineshape of the resonance.}
\end{figure}

In Fig.\,\ref{fig:results} the mean linewidths of pre and post radiated
samples $(\text{\ensuremath{\kappa}}_{\text{after}}+\text{\ensuremath{\kappa}}_{\text{before}})/4\pi$
as well as the difference in linewidths ($\Delta\kappa/2\pi=\text{\ensuremath{\kappa}}_{\text{{after}}}/2\pi-\text{\ensuremath{\kappa}}_{\text{{before}}}/2\pi$)
between the measurements are shown. To determine the possible effects
of the radiation on the Q-factor $\Delta\kappa$ is plotted in histograms
and fitted for each resonator independently with a Gaussian distribution
($P(\Delta\kappa)=A\cdot e^{-(\Delta\kappa-\langle\Delta\kappa_{f}\rangle)^{2}/2\sigma^{2}}$)
in order to localize the center ($\langle\Delta\kappa_{f}\rangle$)
as shown in Fig.\,\ref{fig:results}\,(b). The fits show that the
deviation of the center from 0 is of the order of 1\,\% of the linewidth.
For two of the fits (green and red) the 0 is within the 95\% confidence
interval of the fit for one (dark blue) the full confidence interval
is below 0 and for the last (light blue) it is above 0. This shows
that there is no significant lasting effect of the radiation dose
used. The widths of the distributions from the fits vary between 10
and 20\,MHz. This spread of the distribution is caused by the dependency
of the exact shape of the resonance on the polarization and interference
effects which cause distortions of the lineshape. Examples of data
and fits are shown in Fig.\,\ref{fig:resonanceFits}. The measurements
in Fig.\,\ref{fig:resonanceFits}(a) look very similar but show a
significant deviation in the result of the fit. Although the other
measurements shown almost fall on top of each other for large parts,
the fitted linewidths still deviate by some MHz. The measurement in
Fig.\,\ref{fig:resonanceFits}(d) shows a split resonance which is
fitted with an appropriate function as described above. 

In conclusion we have shown that high-energy proton radiation does
not lead to a degradation of silicon nitride microresonators for Q-factors
of the order of $10^{6}$. Their quality factor did not change due
to the irradiation. This result can be applied to silicon nitride
waveguides with a structure similar to the ones measured here. Our
work therefore paves the way for the rather young platform of silicon
nitride waveguides towards more integrated and robust devices in space
applications. Such devices can be microresonator-based optical frequency
combs.

\subsection*{Acknowledgments}

{\footnotesize{}We thank H. Luckmann and A. Nevsky for assistance
and the PSI Villigen for the irradiation. This work was partially
supported by the European Space Agency (ESA), the DARPA QuSAR program,
by the Bundesministerium für Wirtschaft und Technologie (Germany)
under project no. 50OY1201 and by the Swiss National Science Foundation
(SNF)}{\footnotesize \par}

\bibliographystyle{osajnl}
\bibliography{SiNRad_Arxiv}

\end{document}